\documentclass[sigconf,nonacm]{acmart}
\AtBeginDocument{%
  }

\usepackage{caption}
\usepackage{enumitem}
\usepackage{booktabs}
\usepackage{multirow}
\usepackage[linesnumbered, ruled]{algorithm2e}
\begin{document}

\title{GateDiffInt: Gate-Mediated Controllable Diffusion and Multi-Intent LLM Distillation for User Behavior Modeling}

\author{Jialong Duan}
\orcid{0009-0002-5959-4993}
\authornote{Zichen Zhang and Jialong Duan contributed equally to this work.}
\affiliation{%
  \institution{Fudan University}
  \city{Shanghai}
  \country{China}
}
\email{duanjialong@xiaohongshu.com}

\author{Zichen Zhang}
\orcid{0009-0004-7249-1221}
\authornotemark[1]
\correspondingauthor
\affiliation{%
  \institution{Xiaohongshu Inc.}
  \city{Shanghai}
  \country{China}
}
\email{zhangzichen1@xiaohongshu.com}

\author{Zirui Tu}
\affiliation{%
  \institution{Xiaohongshu Inc.}
  \city{Shanghai}
  \country{China}
}
\email{tuzirui@xiaohongshu.com}

\author{Zheng Zhang}
\affiliation{%
  \institution{Xiaohongshu Inc.}
  \city{Shanghai}
  \country{China}
}
\email{zhangzheng104@xiaohongshu.com}

\author{Zepeng Li}
\affiliation{%
  \institution{Xiaohongshu Inc.}
  \city{Shanghai}
  \country{China}
}
\email{lizepeng@xiaohongshu.com}

\author{Qingyao Cui}
\affiliation{%
  \institution{Xiaohongshu Inc.}
  \city{Beijing}
  \country{China}
}
\email{cuiqingyao@xiaohongshu.com}

\author{Qinwen Wang}
\affiliation{%
  \institution{Xiaohongshu Inc.}
  \city{Beijing}
  \country{China}
}
\email{wangqinwen@xiaohongshu.com}

\author{Yudan Liu}
\correspondingauthor
\affiliation{%
  \institution{Xiaohongshu Inc.}
  \city{Beijing}
  \country{China}
}
\email{liuxiaobo@xiaohongshu.com}

\author{Luo Yang}
\affiliation{%
  \institution{Xiaohongshu Inc.}
  \city{Beijing}
  \country{China}
}
\email{yangluo1@xiaohongshu.com}

\author{Yao Hu}
\affiliation{%
  \institution{Xiaohongshu Inc.}
  \city{Beijing}
  \country{China}
}
\email{xiahou@xiaohongshu.com}

\begin{abstract}
Existing recommendation ranking models, whether traditional feature-interaction methods or sequential approaches with behavior-sequence modeling, typically encode intent signals implicitly in model parameters or hidden states, making it difficult to explicitly disentangle structured intents that vary in strength and temporal scale. More fundamentally, noise and intent in behavior sequences are not independent but mutually reinforcing: on the one hand, behavioral noise persistently dilutes and distorts genuine intents; on the other, the absence of a structured intent prior leaves sequence denoising without a well-defined target. We term this mutual reinforcement Noise--Intent Coupling (NIC). To address NIC, we propose GateDiffInt, an intent interaction framework specifically designed for the ranking stage in industrial recommender systems that uses the final conversion task as a shared signal to jointly align sequence denoising and intent extraction. GateDiffInt introduces a controllable forward diffusion process with dual gating to enhance and denoise user behavior sequences. Building upon the denoised representations, the framework employs a large language model as a teacher to explicitly distill four categories of structured intents---long-term, short-term, latent, and conversion intents---from the behavior sequences, which are then aligned through a lightweight student model. The resulting enhanced sequence representations and structured intent representations are deeply fused via an attention mechanism to produce intent-aware sequence representations, which are subsequently used for final conversion rate prediction. Extensive experiments on both public benchmark datasets and large-scale industrial datasets demonstrate that GateDiffInt consistently and substantially outperforms strong baseline models. Moreover, in large-scale online A/B testing within real-world industrial recommendation scenarios, the ranking model powered by GateDiffInt achieves substantial improvements in GMV and has been successfully deployed to serve the primary traffic for hundreds of millions of daily active users, validating both the effectiveness and practical deployability of the proposed framework in production systems.
\end{abstract}

\maketitle
\section{Introduction}

\begin{figure*}[t]
\centering
\includegraphics[width=\textwidth,height=0.45\textheight,keepaspectratio]{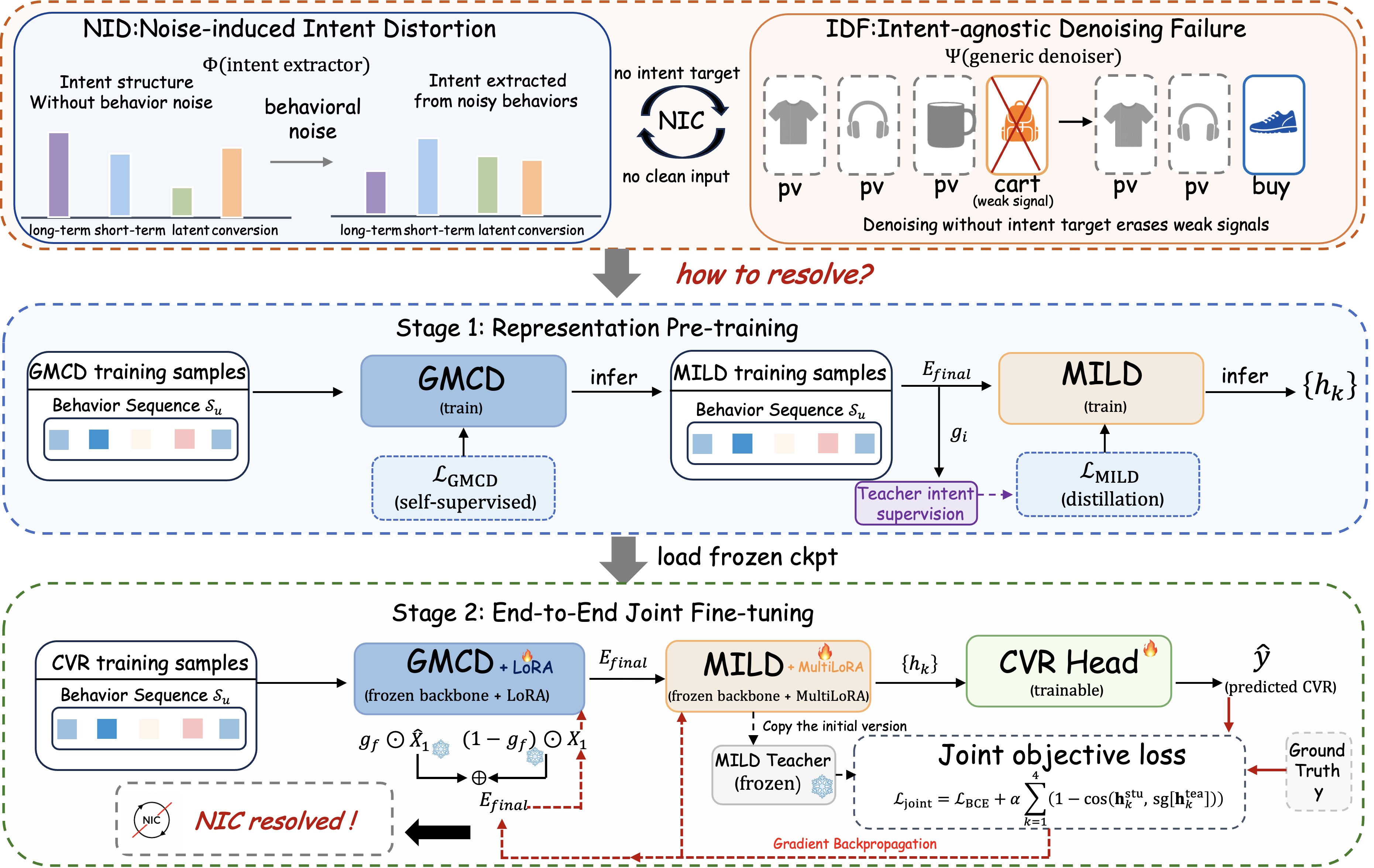}
\caption{Overview of GateDiffInt. Top: the two sub-phenomena of NIC (NID and IDF). Stage 1 pretrains GMCD and MILD separately; Stage 2 loads frozen checkpoints with LoRA and a trainable CVR head, optimizing $ \mathcal{L}_{\rm joint} $ to align denoising and intent extraction toward the conversion task, resolving NIC.}
\label{fig:two_stages}
\end{figure*}

Conversion rate (CVR) prediction operates on the behavior sequences that users generate over a feed, and such sequences inherently carry two entangled ingredients: \emph{noise} from random browsing, accidental clicks, and conformity-driven interactions, and \emph{intents} that evolve in parallel across multiple temporal scales---stable long-term preferences, transient short-term needs, latent comparison, and imminent conversion~\cite{37ESMM,38ESCM2}. Most prior studies treat denoising and intent extraction as separable subproblems: one line extracts intents directly from noisy sequences and leaves the noise to be implicitly absorbed by ever-stronger encoders; another denoises sequences under generic reconstruction targets and defers intent to downstream tasks. Both overlook a more fundamental property: in CVR sequences, noise and intent are not independent but \emph{mutually reinforcing}---noise persistently dilutes and distorts genuine intents, while the absence of a structured intent prior deprives denoising of any criterion for separating informative weak signals from dispensable noise. We term this phenomenon \emph{Noise--Intent Coupling} (NIC): intuitively, intent extraction without clean input has no material, and denoising without an intent target has no aim.

\begin{figure*}[t]
\centering
\includegraphics[width=\textwidth,height=0.4\textheight,keepaspectratio]{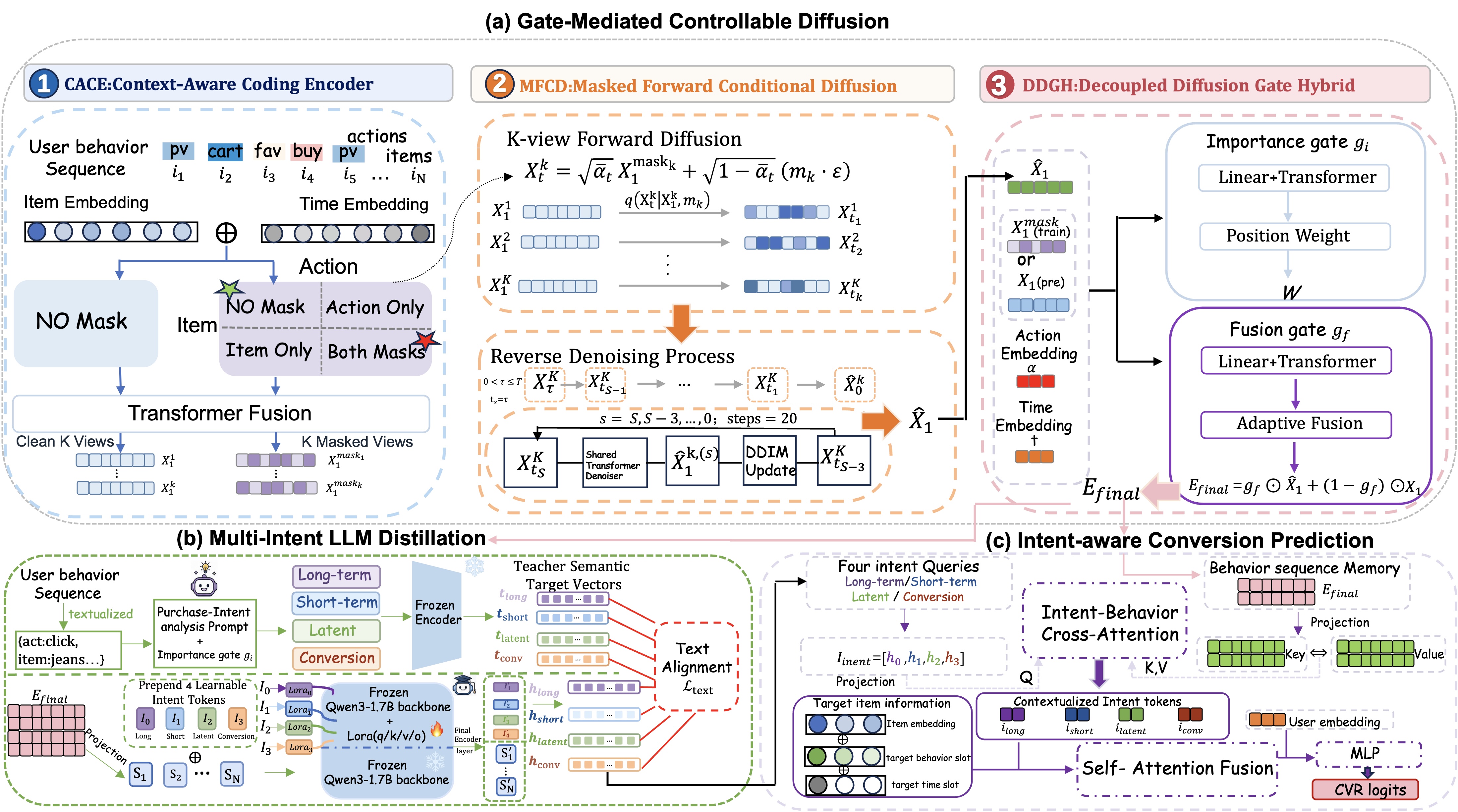}
\caption{Overall architecture of GateDiffInt: (a) Gate-Mediated Controllable Diffusion Module, (b) Multi-Intent LLM Distillation, and (c) Intent-aware Conversion Prediction.}
\label{fig:framework}
\end{figure*}

For analysis and naming, we give NIC a lightweight, \emph{informal} characterization that assumes no specific noise-generating model. Let $\mathcal{I}^\star$ denote a user's underlying intent structure, and \emph{conceptually} decompose the observed sequence as $\mathcal{S}=\mathcal{S}^\star\!\oplus\!\mathcal{N}$ (a true signal $\mathcal{S}^\star$ and a noise component $\mathcal{N}$); let $\Phi$ and $\Psi$ denote the intent extractor and the denoiser, respectively. This characterization serves only to \emph{locate} the two sub-phenomena of NIC, not to impose a strict modeling assumption. NIC manifests as a \emph{mutual conditional degradation} of $\Phi$ and $\Psi$, which decomposes into two separately measurable sub-phenomena. \textbf{\emph{(i) Noise-induced Intent Distortion (NID).}} As $\mathcal{N}$ grows, both attention-based~\cite{13SASRec,14BERT4Rec} and LLM-based~\cite{31LLaRA,32RecFormer} extractors let $\Phi(\mathcal{S})$ drift systematically toward high-frequency yet shallow signals, pulling the output away from $\mathcal{I}^\star$---short-term and latent intents dominate while stable long-term preferences are diluted~\cite{9DIN,42CDR,53KRUGLANSKI}. \textbf{\emph{(ii) Intent-agnostic Denoising Failure (IDF).}} Mainstream diffusion-based recommenders~\cite{20DiffRec,21DiffuRec} optimize generic reconstruction by design and are not conditioned on $\mathcal{I}^\star$, and thus lack a criterion for what to keep: a weak but informative signal such as an early price comparison may be smoothed away together with incidental browsing, or conversely systematic noise may be retained. Even recent diffusion--multi-interest hybrids~\cite{49DiffuMIN,50InDiRec}, though touching both sides, connect $\Phi$ and $\Psi$ through a single forward pass and provide no explicit channel for the two to suppress each other. NID and IDF are therefore not independent problems to be solved in sequence---any one-sided effort leaves a backlash on the other.

Revisiting existing work through the lens of NID and IDF, each line of research closes only one end of the loop. Sequential and multi-interest models~\cite{9DIN,10DIEN,11DSIN,12BST,15MIND,16ComiRec,17SINE,19DMIN} strengthen the representation of the raw sequence through attention or multi-vector modeling, confronting NID head-on; diffusion-based recommenders~\cite{20DiffRec,21DiffuRec,22PreferDiff,23PDRec,24DiffAug,43Diff4Rec,26GCDRec,27DCDR} remove noise under generic reconstruction targets, confronting IDF head-on; recent LLM-based recommenders~\cite{28P5,30TALLRec,31LLaRA,32RecFormer} inject stronger semantic priors into intent extraction, yet still consume unpurified raw sequences; and even diffusion--multi-interest hybrids~\cite{49DiffuMIN,50InDiRec} merely chain the two sides in a single forward pass, lacking a bidirectional mediator. Overall, no existing framework offers a task-grounded shared signal by which denoising and intent extraction iteratively shape each other toward the final objective.

\begin{figure}[h]
    \centering
    \includegraphics[width=\linewidth]{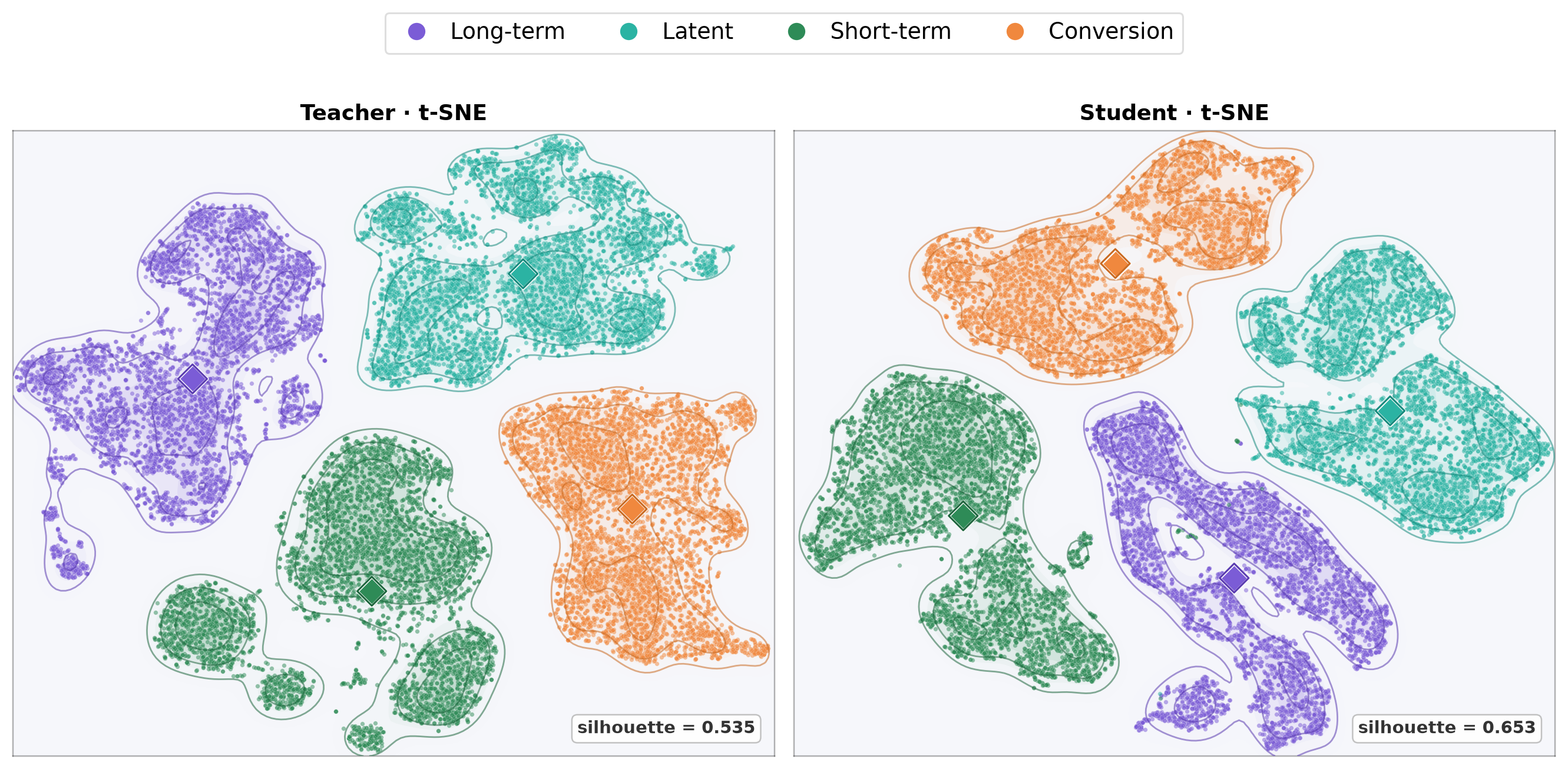}
    \caption{t-SNE of the four distilled intent vectors (long-term, short-term, latent, conversion) on Taobao; the well-separated clusters qualitatively suggest intent disentanglement.}
    \label{fig:tsne}
\end{figure}

Guided by this diagnosis, we propose \textbf{GateDiffInt}, a joint framework that uses the supervisory signal of the final conversion task as a shared guide to align the denoiser $\Psi$ and the intent extractor $\Phi$ under a single objective. The framework comprises three cooperating modules (Figure~\ref{fig:framework}).
 \textbf{\emph{(i) Gate-Mediated Controllable Diffusion (GMCD)}} applies behavior-importance-differentiated masks and forward noise across positions, and performs denoising via mask-aware forward diffusion and deterministic DDIM reverse sampling~\cite{47DDIM,48DDPM}: a \emph{fusion gate} performs position-wise fusion of the denoised and original representations to yield the sequence representation $E_{\rm final}$, while an interpretable position-importance signal is exposed as a reference for downstream intent extraction. \textbf{\emph{(ii) Multi-Intent LLM Distillation (MILD)}} distills a frozen LLM teacher into a lightweight student that produces four structured intent vectors (long-term, short-term, latent, and conversion)---whose categories follow Goal Systems Theory~\cite{53KRUGLANSKI,54Kopetz,55Sugher}---and employs per-intent LoRA routing~\cite{61lora} to keep the four heads from collapsing into a single intent, preserving their distinguishability, which we qualitatively illustrate via a t-SNE visualization (Figure~\ref{fig:tsne}). \textbf{\emph{(iii) Intent-aware Conversion Prediction}} fuses the denoised sequence with the four intents through cross-attention to produce the final conversion prediction.

\begin{figure}[h]
\centering
\includegraphics[width=\linewidth,height=0.08\textheight,keepaspectratio]{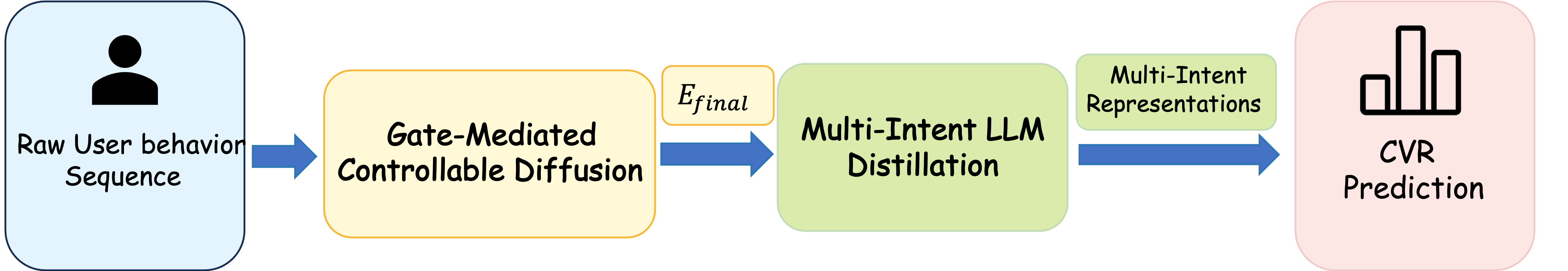}
\caption{Overall pipeline of GateDiffInt.}
\label{fig:pipeline}
\end{figure}

Resolving NIC hinges on a two-stage optimization~(Figure~\ref{fig:two_stages}). \emph{In the first stage}, GMCD and MILD are pretrained separately: GMCD, driven by behavior-importance-differentiated masks and forward noise, learns to preserve reliable signals and suppress noise (mitigating IDF), while MILD distills the four structured intents on top of the enhanced representations (mitigating NID). \emph{In the second stage}, a CVR head is attached for joint fine-tuning: the denoising backbone is frozen and only lightweight LoRA is injected into the sequence encoder and the intent extractor, so that the task-specific supervisory signal refines both at once---the encoder, while preserving sequence reconstruction and enhancement, further down-weights unimportant positions, and intent extraction is drawn closer to the conversion objective---thereby aligning the enhanced sequence representation and the structured intents toward a shared objective and closing the mutual-shaping loop between $\Phi$ and $\Psi$. By using a frozen LLM as the intent teacher and obtaining a lightweight student through distillation, MILD injects the world-knowledge priors of LLMs while meeting the online-latency constraints of industrial ranking, which also distinguishes it from paradigms that employ LLMs directly as recommenders~\cite{28P5,30TALLRec,31LLaRA}. The main contributions of this work are as follows:

\begin{itemize}[leftmargin=*,itemsep=3pt,parsep=0pt,topsep=3pt]
\item \textbf{We identify and empirically characterize \emph{Noise--Intent Coupling} (NIC) in CVR sequence modeling.} To our knowledge, we are the first to explicitly identify the mutual reinforcement in which noise degrades intent and unstructured intent degrades denoising, decomposing it into two separately measurable sub-phenomena, NID and IDF, and empirically confirming their existence and magnitude through a diagnostic study (Section~\ref{sec:diagnostic}).

\item \textbf{We propose GateDiffInt, which resolves NIC via a task-grounded shared signal.} Through a two-stage design, the supervisory signal of the final conversion task serves as a shared guide connecting denoising and intent extraction, refining the sequence encoding and the intent extractor simultaneously via lightweight LoRA so that the two align toward a shared objective.

\item \textbf{We validate GateDiffInt at both benchmark and industrial scale.} On two public benchmarks and a production dataset from an industrial recommender serving hundreds of millions of daily active users, GateDiffInt consistently outperforms representative sequential and multi-interest ranking baselines under a unified CVR ranking task; component-wise ablations further isolate the individual gains of controllable diffusion and multi-intent LLM distillation, a 14-day online A/B test corroborates the offline improvements, and the framework has been deployed at scale.
\end{itemize}
\section{Related Work}
We review related work along four lines and position GateDiffInt through the lens of NID and IDF.

\textbf{Sequential and multi-interest models.} Beyond feature-interaction models~\cite{1DeepFM,2DCN,3DCNV2,4xDeepFM,5AutoInt,6FiBiNET,7AFN,8FinalMLP}, DIN, DIEN, DSIN, BST, and HSTU~\cite{9DIN,10DIEN,11DSIN,12BST,25HSTU} model sequential dynamics via attention, interest evolution, and Transformers, while MIND, ComiRec, and DMIN~\cite{15MIND,16ComiRec,17SINE,18PinnerSage,19DMIN,44DMRL,45MSR} further capture interest diversity with multiple latent vectors. However, all of them operate directly on the raw noisy sequence, implicitly assuming that noise can be absorbed by the encoder~\cite{42CDR}, and their interests emerge at the co-occurrence level without alignment to a semantic structure---hence they are prone to NID under noise. In contrast, GateDiffInt first explicitly purifies the sequence via controllable diffusion and then distills semantically labeled structured intents from an LLM.

\textbf{Diffusion-based recommenders.} DiffRec and DiffuRec~\cite{20DiffRec,21DiffuRec} pioneer the use of diffusion for sequence denoising, followed by guided, curriculum-based, and conditional variants~\cite{22PreferDiff,23PDRec,24DiffAug,43Diff4Rec,26GCDRec,27DCDR}. Yet their forward noising and reconstruction targets are largely position-agnostic and lack fine-grained control aligned with behavioral importance and intent structure; absent an intent prior, they cannot decide which weak signals to keep---i.e., they suffer from IDF. GMCD instead applies behavior-reliability-differentiated masks and forward noise, and fuses the denoised and original representations position-wise via gating, thereby preserving latent-intent weak signals while suppressing noise.

\textbf{LLMs for recommendation and intent modeling.} LLMs have recently been used to unify recommendation paradigms or to enhance sequential models~\cite{28P5,30TALLRec,31LLaRA,32RecFormer}, but mainstream practice targets generative recommendation or post-hoc explanation~\cite{33LLM4Rec-Survey1,34LLM4Rec-Survey2}, incurring high online latency, still consuming raw sequences, and rarely producing structured intents directly consumable by ranking models. Knowledge distillation~\cite{35KD-SR,36ActiveKD,41KD,56KDsurvey} can transfer such capabilities but has not been organically combined with sequence denoising. MILD positions the LLM as an offline structured-intent teacher and distills it into a lightweight student via per-intent LoRA, injecting world-knowledge priors while meeting online-latency constraints---fundamentally different from employing LLMs directly as recommenders.

\begin{figure*}[t]
\centering
\includegraphics[width=\textwidth,height=0.15\textheight,keepaspectratio]{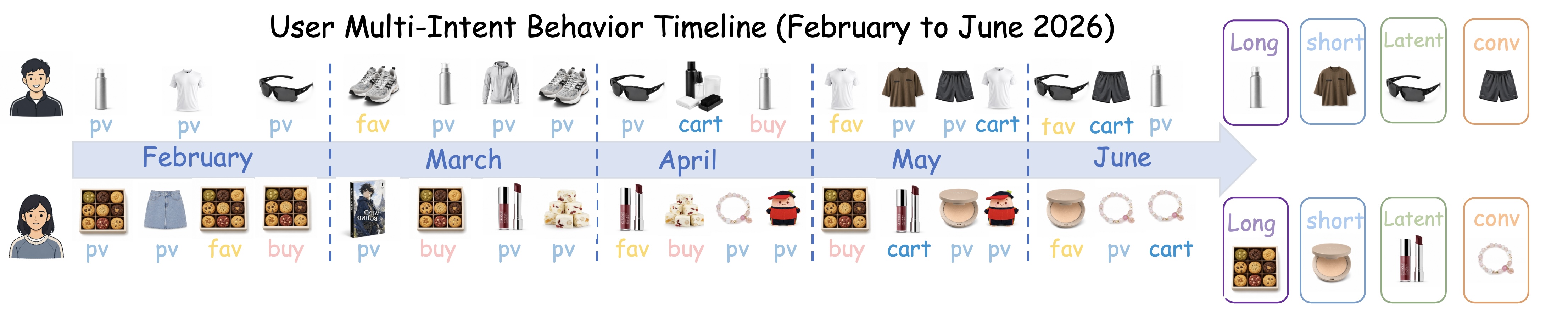}
\caption{Multi-intent behavior timelines of two e-commerce users (Feb--Jun 2026). Right panels show items linked to long-term, short-term, latent, and conversion intents.}
\label{fig:Four_intents}
\end{figure*}

\textbf{Diffusion--multi-interest hybrids.} DiffuMIN and InDiRec~\cite{49DiffuMIN,50InDiRec} combine diffusion with multi-interest modeling and thus touch both the denoising and intent sides, but connect them through a one-way denoise-then-extract pass, lacking a bidirectional mediator for the two to correct each other and thus failing to close the NIC loop. GateDiffInt instead uses the supervisory signal of the final conversion task as a shared guide and, through two-stage optimization, aligns denoising and intent extraction toward a common objective, mechanistically closing the mutual-shaping loop between $\Phi$ and $\Psi$.
\section{Methodology}
In this section, we briefly review the background and then detail the proposed GateDiffInt framework and its key components. 

Let $\mathcal{U}$ and $\mathcal{V}$ denote the sets of users and items, respectively, and let $\mathcal{A}$ be the set of interaction types. For each user $u \in \mathcal{U}$, we take the most recent $L$ interactions in chronological order to form the behavioral sequence
\begin{equation}
\mathcal{S}_u = \bigl\{(i_\ell,\, a_\ell,\, \Delta t_\ell)\bigr\}_{\ell=1}^{L},
\end{equation}
where $i_\ell \in \mathcal{V}$ is the item, $a_\ell \in \mathcal{A}$ the action type, and $\Delta t_\ell \in \mathbb{R}_{\ge 0}$ the time interval to the current timestamp; sequences shorter than $L$ are left-padded. Given a candidate item $v^{\rm tgt} \in \mathcal{V}$, conversion-rate (CVR) prediction estimates the purchase probability
\begin{equation}
\hat{y} = f_\Theta\bigl(\mathcal{S}_u,\, v^{\rm tgt},\, u\bigr) \in [0,1],
\end{equation}
where $f_\Theta$ is the model parameterized by $\Theta$, optimized by the binary cross-entropy between $\hat{y}$ and the ground-truth label $y \in \{0,1\}$. Since the behavioral sequence entangles noise with multi-scale intents, GateDiffInt comprises three cooperating modules: \textbf{GMCD} performs controllable denoising and passes a fused representation $E_{\rm final}$ together with an interpretable importance signal $g_i$ to the downstream MILD; \textbf{MILD} distills four structured intent vectors $\{\mathbf{h}_k\}_{k=1}^{4}$ (long-term, short-term, latent, and conversion) from $E_{\rm final}$; and an \textbf{Intent-aware Conversion Prediction} module fuses them to produce the prediction. The overall data flow is $\mathcal{S}_u \to (E_{\rm final},\, g_i) \to \{\mathbf{h}_k\} \to \hat{y}$.

\subsection{GMCD}
User behavioral sequences can be viewed as noisy observations of true preferences: shallow actions are noise-heavy, whereas deep actions, though sparse, carry strong signals. GMCD performs controllable denoising in the latent space and outputs a position-wise aligned fused representation $E_{\rm final}\in\mathbb{R}^{L\times d}$ as the primary sequence representation for the downstream MILD, together with an interpretable importance signal $g_i$ that the MILD teacher can consult. GMCD consists of three cascaded sub-modules:
\begin{itemize}[leftmargin=*,itemsep=2pt,parsep=0pt,topsep=2pt]
    \item \textbf{Context-Aware Coding Encoder (CACE)}: fuses item and temporal information and applies action-aware masking to produce a clean view $X_1$ and a masked view $X_1^{\rm mask}$;
    \item \textbf{Masked Forward Conditional Diffusion (MFCD)}: performs mask-aware forward diffusion and partial-start deterministic reverse sampling to obtain the denoised reconstruction $\hat{X}_1$;
    \item \textbf{Decoupled Diffusion Gate Hybrid (DDGH)}: fuses $\hat{X}_1$ and $X_1$ via a fusion gate $g_f$ into $E_{\rm final}$, and emits an interpretable importance signal via an importance gate $g_i$ for the downstream MILD teacher.
\end{itemize}

Training builds self-supervision from masking and noise injection, while inference uses deterministic sampling for efficiency.

\subsubsection{CACE}
For each position $\ell$, the item and temporal embeddings are linearly projected, concatenated with a learnable fusion token, and fed into a position-wise shared Transformer that attends only within each position (never across positions), decoupling feature fusion from sequential modeling. Its output is the fused representation $\mathbf{z}_\ell$, and the clean view is
\begin{equation}
X_1=[\mathbf{z}_1,\dots,\mathbf{z}_L]\in\mathbb{R}^{L\times d}.
\end{equation}
For every non-padding position, CACE independently samples two Bernoulli masks: an item mask for item-level reconstruction and an action mask set by behavioral reliability---higher for shallow actions (e.g., \texttt{pv}) and lower for strong actions (e.g., \texttt{buy}):
\begin{equation}
M^{\rm item}_\ell\sim\mathrm{Bernoulli}(p^{\rm item}),
M^{\rm act}_\ell\sim\mathrm{Bernoulli}\bigl(\pi[a_\ell]\bigr).
\end{equation}
Masked items are replaced by a learnable \texttt{<MASK>} token and re-encoded (temporal information retained), yielding the masked view $X_1^{\rm mask}$. Padding positions are never masked, and the two masks jointly determine the per-position noise strength in the subsequent forward diffusion. We sample $K{=}2$ masked views per sequence during training; at inference no masking is applied and $X_1$ is used as the reverse-sampling start.

\subsubsection{MFCD}
MFCD runs diffusion in the CACE latent space, sharing one denoising network $f_\theta$ across training and inference. Under the DDPM~\cite{48DDPM} parameterization with a cosine schedule and $X_1$ as the reconstruction target, each masked view is noised as
\begin{equation}
X_t^{k}=\sqrt{\bar\alpha_t}\,X_1^{\rm mask_k}+\sqrt{1-\bar\alpha_t}\cdot\bigl(m_k(\ell)\,\varepsilon\bigr),\varepsilon\sim\mathcal{N}(0,I),
\end{equation}
where the noise multiplier $m_k(\ell)$ is set by the mask state at $\ell$ (baseline for no mask, strongest for dual mask), so more-masked positions receive stronger noise. The timestep- and position-conditioned $f_\theta$ predicts $x_0$, giving the reconstruction loss
\begin{equation}
\mathcal{L}_{\rm denoise}=\frac{1}{K}\sum_{k=1}^{K}\bigl\|\hat{X}_1^{k}-\mathrm{sg}[X_1]\bigr\|_2^{2}.
\end{equation}
An InfoNCE regularizer over the denoised outputs of two masked views enhances sequence-level discriminability:
\begin{equation}
\mathcal{L}_{\rm cl}=-\frac{1}{K(K-1)B}\sum_{b=1}^{B}\sum_{\substack{p,q=1 \\ p\neq q}}^{K}\log\frac{\exp\bigl(\mathrm{sim}(\mathbf{z}_b^{(p)},\mathbf{z}_b^{(q)})/\tau\bigr)}{\sum_{b'=1}^{B}\exp\bigl(\mathrm{sim}(\mathbf{z}_b^{(p)},\mathbf{z}_{b'}^{(q)})/\tau\bigr)},
\end{equation}
where $\mathbf{z}^{(k)}=\phi(\mathrm{Pool}(\hat{X}_1^{k}))$ and $\mathrm{sim}$ is cosine similarity. At inference, starting from an intermediate step $s=\lfloor\rho T\rfloor$,
\begin{equation}
X_s=\sqrt{\bar\alpha_s}\,X_1+\sqrt{1-\bar\alpha_s}\,\varepsilon,
\end{equation}
a deterministic DDIM~\cite{47DDIM} update ($\eta=0$) is applied along $t_1>t_2>\cdots>t_{K'}=0$ (with $t_1=s$):
\begin{equation}
\hat\varepsilon_{t_i}=\frac{X_{t_i}-\sqrt{\bar\alpha_{t_i}}\,\hat{X}_1^{(t_i)}}{\sqrt{1-\bar\alpha_{t_i}}},
X_{t_{i+1}}=\sqrt{\bar\alpha_{t_{i+1}}}\,\hat{X}_1^{(t_i)}+\sqrt{1-\bar\alpha_{t_{i+1}}}\,\hat\varepsilon_{t_i}.
\end{equation}
The resulting $\hat{X}_1$ is passed together with $X_1$ to DDGH.

\subsubsection{DDGH}
DDGH fuses $\hat{X}_1$ and $X_1$ via dual gates that share the same architecture but have independent parameters: $[\hat{X}_1;X_1^{\rm mask};\mathbf{e}^{a};\mathbf{e}^{t}]$ is concatenated and passed through a lightweight Transformer to produce gate values,
\begin{equation}
g=\sigma\Big(\mathrm{MLP}\big(\mathrm{Trans}_{\rm gate}([\hat{X}_1;\,X_1^{\rm mask};\,\mathbf{e}^{a};\,\mathbf{e}^{t}])\big)\Big)\in[0,1]^{L}.
\end{equation}
The fusion gate $g_f$ performs position-wise weighted fusion,
\begin{equation}
E_{\rm final}=g_f\odot\hat{X}_1+(1-g_f)\odot X_1,
\end{equation}
while the importance gate $g_i$ does not participate in the fusion; instead it produces an interpretable position-importance signal provided to the downstream MILD teacher as a reference. Both gates are supervised by self-generated targets with BCE. Let the per-position denoising error be $u_\ell=\|\hat{X}_{1,\ell}-X_{1,\ell}\|_2^2/d$ with sequence-wise normalized form $\tilde{u}$; $g_f$ is supervised by $1-\tilde{u}$, and $g_i$ further incorporates a behavioral prior $b_\ell$ ($\lambda$ balances the prior and the reconstruction error):
\begin{equation}
t^{(i)}_\ell=\mathrm{clip}\big(\lambda b_\ell+(1-\lambda)\tilde{u}_\ell,\ [t_{\min},t_{\max}]\big),
\end{equation}
\begin{equation}
\mathcal{L}_{\rm fusion}=\mathrm{BCE}\big(g_f,\,\mathrm{sg}[1-\tilde{u}]\big),
\mathcal{L}_{\rm imp}=\mathrm{BCE}\big(g_i,\,\mathrm{sg}[t^{(i)}]\big).
\end{equation}
The overall GMCD objective sums the denoising, contrastive, and dual-gate losses, where $\lambda_{\rm cl}$, $\lambda_f$, and $\lambda_i$ are balancing coefficients:
\begin{equation}
\mathcal{L}_{\rm GMCD}=\mathcal{L}_{\rm denoise}+\lambda_{\rm cl}\mathcal{L}_{\rm cl}+\lambda_f\mathcal{L}_{\rm fusion}+\lambda_i\mathcal{L}_{\rm imp}.
\end{equation}

GMCD thus completes encoding, diffusion-based denoising, and gated fusion, passing $E_{\rm final}$ and the importance signal $g_i$ to MILD.

\subsection{MILD}\label{subsec:mild}
Users pursue multiple goals of different temporal scales and priorities. Following Goal Systems Theory~\cite{53KRUGLANSKI} and subsequent consumer-behavior studies~\cite{54Kopetz,55Sugher}, we distinguish four categories of intent (Figure~\ref{fig:Four_intents}):

\begin{itemize}[leftmargin=*,itemsep=2pt,parsep=0pt,topsep=2pt]
    \item \textbf{Long-term intent}: rooted in identity and lifestyle, stable across contexts;
    \item \textbf{Short-term intent}: triggered by events or state changes within a clear time window;
    \item \textbf{Latent intent}: reflected by repeated browsing and comparison without conversion;
    \item \textbf{Conversion intent}: the most probable immediate buy decision given prior signals and current context.
\end{itemize}
On the teacher side, MILD employs a frozen Gemini~3.5~Flash (via API)~\cite{62gemini} to generate textual descriptions of the four intents from a user's behavioral sequence as semantic supervision; the importance signal $g_i$ from GMCD is supplied to the teacher as a position-importance hint alongside the sequence, so that the teacher attends more to reliable behaviors when generating the descriptions. On the student side, a frozen Qwen3-1.7B backbone~\cite{59Qwen3} extracts the four intent vectors via per-intent LoRA~\cite{61lora} routing, and Qwen3-Embedding-8B~\cite{60Qwen3Emb} encodes the teacher texts into the semantic space for alignment.

Concretely, $K{=}4$ learnable intent tokens $\{I_0,\dots,I_{K-1}\}$ are prepended, and $E_{\rm final}$ is projected into the LLM latent space and concatenated:
\begin{equation}
\mathbf{H}^{(0)}=[I_0,\dots,I_{K-1},\;\mathrm{Proj}(E_{\rm final})],
\end{equation}
where $\mathrm{Proj}$ maps from dimension $d$ to $d_{\rm LLM}$. A 2D attention mask lets each intent token attend only to itself and all behavior positions (intent tokens are mutually invisible), while behavior positions follow a causal mask and cannot attend to any intent token. During adaptation, LoRA is injected only into the $\{q,k,v,o\}$ projections; the $k$-th intent token activates only the $k$-th independent LoRA path, and behavior tokens receive no LoRA updates. The hidden states at the first $K$ positions are taken as the intent vectors:
\begin{equation}
[\mathbf{h}_{\rm long},\,\mathbf{h}_{\rm short},\,\mathbf{h}_{\rm latent},\,\mathbf{h}_{\rm conv}]=\mathrm{Backbone}(\mathbf{H}^{(0)},M)[:,:K,:],
\end{equation}
where $M$ is the attention mask above.

Teacher descriptions are encoded by Qwen3-Embedding-8B into targets $\mathbf{t}_{b,k}^{\rm tea}$; student intents are mapped by a projection head $\mathrm{Head}(\cdot)$ into the teacher embedding space and aligned via a masked cosine distance:
\begin{equation}
\mathcal{L}_{\rm MILD}=\frac{1}{\sum_{b,k}m_{b,k}}\sum_{b,k}m_{b,k}\Big(1-\cos\big(\mathrm{Head}(\mathbf{h}_{b,k}),\,\mathbf{t}_{b,k}^{\rm tea}\big)\Big),
\label{eq:mild}
\end{equation}
where $b$ indexes samples in a batch, $k$ indexes intent categories, and $m_{b,k}\in\{0,1\}$ indicates whether valid supervision is available; unsupervised positions are excluded from the loss.

MILD thus extracts the four structured intent vectors via per-intent LoRA routing and semantic distillation, and passes them together with $E_{\rm final}$ to the downstream intent-aware conversion prediction module.

\subsection{Intent-aware Conversion Prediction}
Given $E_{\rm final}$ and the four intent vectors $\{\mathbf{h}_k\}_{k=1}^{4}$, this module fuses them with candidate item information to produce the conversion probability $\hat{y}$.

Cross-attention is performed with intents as queries and the behavioral sequence as keys and values. On the sequence side, $E_{\rm final}$ is concatenated with behavior embeddings and normalized; on the intent side, a linear transform with nonlinear activation yields the query, followed by multi-head cross-attention to obtain contextualized intents:
\begin{equation}
\mathbf{e}^{\rm new}_\ell=\mathrm{LN}\big([W_e E_{\rm final,\ell};\,\mathbf{e}^{\rm beh}_\ell]\big),
\end{equation}
\begin{equation}
\mathbf{i}=\mathrm{LN}\big(\phi(W_i\mathbf{h})\big),
\end{equation}
\begin{equation}
\mathbf{i}^{\rm out}=\mathrm{LN}\big(\mathbf{i}+\mathrm{CrossAttn}(Q=\mathbf{i},\,K=V=\mathbf{e}^{\rm new})\big).
\end{equation}

The contextualized intents $\{\mathbf{i}^{\rm out}_k\}_{k=1}^{4}$ are concatenated with candidate item features and passed through self-attention to obtain the fused representation $\mathbf{T}^{\rm out}$, which is then fed together with the user embedding into an MLP to produce the final prediction under the binary cross-entropy loss $\mathcal{L}_{\rm BCE}$:
\begin{equation}
\hat{y}=\sigma\big(\mathrm{MLP}\big([\mathrm{flatten}(\mathbf{T}^{\rm out});\,\mathbf{e}^{\rm user}]\big)\big).
\end{equation}

GateDiffInt is trained in two stages (Figure~\ref{fig:two_stages}): GMCD and MILD are first pretrained separately; checkpoints are then loaded and a CVR head is attached for end-to-end joint fine-tuning. During fine-tuning, the GMCD backbone is frozen and only a low-rank LoRA is injected into the encoder fusion Transformer; MILD retains its per-intent MultiLoRA; the CVR head is fully trainable. To preserve the distilled intent geometry, a teacher-anchoring regularizer is introduced:
\begin{equation}
\mathcal{L}_{\rm joint}=\mathcal{L}_{\rm BCE}+\alpha\sum_{k=1}^{4}\big(1-\cos(\mathbf{h}_k^{\rm stu},\,\mathrm{sg}[\mathbf{h}_k^{\rm tea}])\big),
\end{equation}
where $\mathbf{h}_k^{\rm tea}$ comes from the frozen MILD copy after the first stage, and $\alpha$ is a balancing coefficient. Since DDIM reverse sampling does not propagate gradients and the fusion gate is frozen at this stage, CVR gradients reach the encoder LoRA only through the $(1-g_f)\odot X_1$ branch, enabling lightweight refinement.

\section{Experiments}
In this section, we conduct extensive experiments to evaluate the effectiveness of GateDiffInt\footnote{Code is available at: \url{https://github.com/chen667/GateDiffInt}}.

\begin{figure*}[t]
\centering
\includegraphics[width=\textwidth]{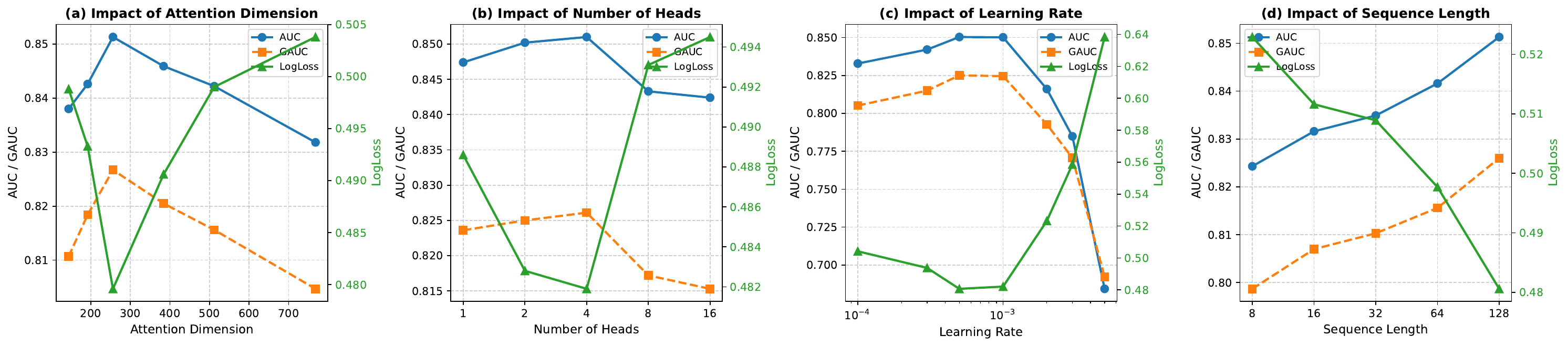}
\caption{Hyper-parameter sensitivity of our model on the Taobao dataset: (a) attention dimension, (b) number of heads, (c) learning rate, and (d) sequence length. Left axis: AUC and GAUC; right axis: LogLoss.}
\label{fig:hyper}
\end{figure*}

\subsection{Experimental Setup}
\subsubsection{Datasets} We evaluate on two public benchmarks and one industrial dataset. \textbf{Taobao}\footnote{\url{https://tianchi.aliyun.com/dataset/dataDetail?dataId=649}} records behavior logs (click, add-to-cart, favorite, and buy) of about one million users from Nov.~25 to Dec.~3, 2017, yielding \textbf{1.67M} training and \textbf{0.53M} test samples over \textbf{5} features; positives are real buy behaviors, and negatives are sampled at a 1:1 ratio, preferentially as hard negatives from interacted-but-unpurchased items. \textbf{Amazon-Electronics}\footnote{\url{https://snap.stanford.edu/data/amazon/productGraph/categoryFiles/reviews_Electronics_5.json.gz}}~\cite{57Amazon} consists of review records, from which item embeddings are obtained directly from review texts via a large language model, giving \textbf{1.56M}/\textbf{0.30M} train/test samples over \textbf{9} features; ratings $\geq 4$ are treated as positive, and negatives are drawn by popularity$^{0.75}$ sampling~\cite{58NegtiveSample} from non-interacted items. \textbf{Industrial} is sampled from the click logs of a large-scale recommender serving hundreds of millions of users, using one full day of logs for training/validation and a subset of the next day for testing, yielding \textbf{20M}/\textbf{2M} train/test samples over \textbf{56} features. To prevent user-level leakage, all datasets reserve disjoint user sets exclusively for GMCD pre-training and MILD distillation, with the remaining users forming the CVR evaluation set under the leave-one-out protocol~\cite{13SASRec,14BERT4Rec}.

\subsubsection{NIC Diagnostic Analysis}
\label{sec:diagnostic}
To show that Noise--Intent Coupling (NIC) is a \emph{measurable} phenomenon
rather than a conceptual abstraction, we isolate its two sub-phenomena---NID
and IDF---on the public Taobao benchmark (3 seeds; mean$\pm$std).
\noindent\textbf{NID (Noise-induced Intent Distortion).}
We inject replacement noise at ratio $\gamma\in\{0,0.1,\dots,0.5\}$ into each test
sequence, replacing a $\gamma$ fraction of non-padding, non-purchase positions with
popularity-sampled random items while protecting the target and strong behaviors.
Using the clean ($\gamma{=}0$) output as reference, we measure the intent drift
\begin{equation}
D(\gamma)=\frac{1}{K}\sum_{k=1}^{K}\Big(1-\cos\big(\mathbf{h}_k(\mathcal{S}_0),\,
\mathbf{h}_k(\mathcal{S}_\gamma)\big)\Big),
\end{equation}
over the $K{=}4$ intent vectors. The denoising-free variant (\emph{w/o GMCD})
drifts sharply and monotonically as noise grows, whereas GateDiffInt stays
markedly stable---its drift is consistently $\sim$$6.5\times$ lower across all
$\gamma$---confirming both the existence of NID and its mitigation by our
task-grounded denoising.
\noindent\textbf{IDF (Intent-agnostic Denoising Failure).}
On a weak-signal subset (cart/fav interactions occurring more than the median interval before the final purchase; $N_w{\approx}1.8{\times}10^{5}$), we compare our controllable diffusion against a \emph{vanilla} diffusion that is identical in encoder and training data but strips all task grounding (plain DDPM, no gating/conditioning). We report the weak-signal reconstruction fidelity $\cos(\hat{X}_{\mathrm{denoised}},X_{\mathrm{clean}})$. Vanilla diffusion attains only $0.44$---its intent-agnostic denoising severely distorts the encoded weak signals---whereas GateDiffInt faithfully preserves them at $0.98$, a $2.2\times$ improvement, confirming IDF and its mitigation. As weak-signal positions are only a small fraction of each sequence, this large fidelity gap is diluted into the moderate AUC gap seen in Table~\ref{tab:ablation}.
These diagnostics empirically ground NIC: noise distorts intent when
denoising is absent (NID), and generic denoising erases informative weak signals when it is not intent-grounded (IDF)---both addressed by GateDiffInt.

\subsubsection{Baselines} We compare GateDiffInt against representative sequential baselines under a unified CVR ranking task:
\begin{itemize}[leftmargin=*,itemsep=2pt,parsep=0pt,topsep=2pt]
    \item \textbf{DIN, DIEN, DSIN}~\cite{9DIN,10DIEN,11DSIN}: target attention and interest evolution over historical behaviors;
    \item \textbf{BST, HSTU}~\cite{12BST,25HSTU}: Transformer self-attention for long-range dependencies;
    \item \textbf{DMIN}~\cite{19DMIN}: multi-vector modeling of interest diversity.
\end{itemize}
We do not include diffusion-based/LLM-based recommenders as standalone baselines, as they target top-$k$ or generative recommendation and are misaligned with target-conditioned pointwise CVR ranking; instead, the benefits of controllable diffusion and multi-intent LLM distillation are isolated via component-wise ablations. All baselines follow the implementation practice of FuxiCTR~\cite{63Fuxictr}, use the same input features and sequence length as ours, and are optimized with BCE; since Taobao has few features, we additionally train an item encoder on it.

\subsubsection{Evaluation Metrics}
We adopt AUC (Area Under the ROC Curve), GAUC (Group AUC), and LogLoss as the offline evaluation metrics, and GMV (Gross Merchandise Volume) as the business metric for online A/B testing. AUC measures overall ranking ability, LogLoss measures the calibration of predicted probabilities, while GAUC further evaluates ranking quality within each user group and has been shown to be more consistent with online performance~\cite{29GAUC1,46GAUC2}. Its calculation is given in Equation~\eqref{eq:gauc}:
\begin{equation}
\text{GAUC} = \frac{\sum_{u=1}^{U} \#samples(u) \times \text{AUC}_u}{\sum_{u=1}^{U} \#samples(u)}.
\label{eq:gauc}
\end{equation}
where \(U\) denotes the number of users, \(\#samples(u)\) is the number of samples for the \(u\)-th user, and \(\text{AUC}_u\) is the AUC computed on the samples of user \(u\).

\begin{table*}[t]
\centering
\caption{Performance comparison on the public datasets. The best results are highlighted in bold, and the second-best results are underlined. Improvements over the strongest baseline are statistically significant with $p$-value $< 0.01$ under pairwise $t$-tests.}
\label{tab:public_results}
\begin{tabular}{lcccccc}
\toprule
\multirow{2}{*}{Model} & \multicolumn{3}{c}{Taobao} & \multicolumn{3}{c}{Amazon-Electronics} \\
\cmidrule(lr){2-4} \cmidrule(lr){5-7}
 & AUC & GAUC & LogLoss & AUC & GAUC & LogLoss \\
\midrule
DIN & 0.8068$\pm$0.0039 & 0.7919$\pm$0.0035 & 0.5353$\pm$0.0039 & 0.7141$\pm$0.0105 & 0.6996$\pm$0.0106 & 0.6228$\pm$0.0091 \\
DIEN & 0.8388$\pm$0.0011 & 0.8152$\pm$0.0010 & 0.4974$\pm$0.0031 & 0.7523$\pm$0.0037 & 0.7399$\pm$0.0037 & 0.6002$\pm$0.0058 \\
DSIN & 0.8254$\pm$0.0073 & 0.8033$\pm$0.0061 & 0.5134$\pm$0.0078 & 0.7612$\pm$0.0021 & 0.7489$\pm$0.0025 & 0.5915$\pm$0.0049 \\
BST & 0.8378$\pm$0.0021 & 0.8126$\pm$0.0012 & 0.5005$\pm$0.0015 & 0.7619$\pm$0.0050 & 0.7491$\pm$0.0048 & 0.5889$\pm$0.0077 \\
DMIN & \underline{0.8397$\pm$0.0013} & \underline{0.8174$\pm$0.0039} & \underline{0.4963$\pm$0.0040} & 0.7636$\pm$0.0033 & 0.7527$\pm$0.0026 & 0.5878$\pm$0.0068 \\
HSTU & 0.8304$\pm$0.0032 & 0.8075$\pm$0.0031 & 0.5064$\pm$0.0043 & \underline{0.7829$\pm$0.0016} & \underline{0.7681$\pm$0.0019} & \underline{0.5693$\pm$0.0047} \\
\textbf{Ours} & \textbf{0.8515$\pm$0.0014} & \textbf{0.8275$\pm$0.0013} & \textbf{0.4836$\pm$0.0018} & \textbf{0.8016$\pm$0.0076} & \textbf{0.7792$\pm$0.0033} & \textbf{0.5414$\pm$0.0046} \\
\bottomrule
\end{tabular}
\end{table*}

\subsubsection{Implementation Details}
All methods use the same backbone architectures as in the original papers. We optimize all models with Adam using a learning rate of \(5 \times 10^{-4}\). Each model is trained for up to 15 epochs with early stopping (patience = 4). The batch size is 2048 for baselines, 96 for GateDiffInt on public datasets, and 64 for GateDiffInt on the production dataset. All models are implemented in PyTorch. The maximum sequence length is 128 for Taobao (average length $\approx$ 101), 32 for Amazon-Electronics (average length $\approx$ 9), and 1024 for the production dataset. For the diffusion process, timesteps are uniformly sampled from 1 to 1000 in the forward process, and the reverse process starts at timestep 700 with DDIM sampling~\cite{47DDIM,48DDPM}.

\subsection{Performance on Public Datasets}
\subsubsection{Overall Performance} Table~\ref{tab:public_results} reports the overall performance on the two public datasets. Sequential models consistently outperform DIN on both Taobao and Amazon-Electronics, with stronger sequential modeling yielding better results. On Taobao, DIEN and DMIN are the strongest baselines, while HSTU performs best on Amazon-Electronics. Notably, HSTU underperforms on Taobao due to limited feature dimensions and short sequence length, but benefits from richer review-based embeddings on Amazon-Electronics. GateDiffInt achieves the best results across all metrics on both datasets, with statistically significant improvements over the strongest baselines, demonstrating the effectiveness of unifying controllable diffusion-based sequence enhancement with multi-intent LLM distillation.

\subsubsection{Impact of Hyperparameters}
For simplicity, we study the sensitivity of GateDiffInt to key hyperparameters on the Taobao dataset, including attention dimension, number of attention heads, learning rate, and sequence length (Figure~\ref{fig:hyper}).

The learning rate is the most sensitive: the model performs best within \(5\times10^{-4}\) to \(1\times10^{-3}\), while lower rates cause underfitting and higher rates (\(\geq 2\times10^{-3}\)) lead to instability. The attention dimension shows an inverted-U curve peaking at 256. The model is relatively robust to the number of heads (best at 4), and performance improves steadily with longer sequences. Overall, the default setting (dimension=256, heads=4, learning rate=\(5\times10^{-4}\)) lies in a favorable range.

\subsection{Performance on Industrial Dataset}
Table~\ref{tab:production_results} reports the results on the production dataset. Baseline performance improves with stronger sequential modeling, and HSTU achieves the best results among all baselines. Unlike its weaker performance on Taobao, HSTU shows a clear advantage here, benefiting from the longer sequences and richer features in the production data. GateDiffInt significantly outperforms all baselines, achieving relative improvements of 3.00\% in AUC, 1.82\% in GAUC, and 2.81\% in LogLoss over HSTU. These consistent gains in both ranking quality and calibration further validate the effectiveness of our approach in real-world industrial scenarios.

\begin{table}[h]
\centering
\caption{Performance comparison on the production dataset. The best results are highlighted in bold. Improvements over the strongest baseline are statistically significant with $p$-value $< 0.01$ under pairwise $t$-tests.}
\label{tab:production_results}
\begin{tabular}{lccc}
\toprule
Model & AUC & GAUC & LogLoss \\
\midrule
DIN   & 0.7682 $\pm$ 0.0026 & 0.7053 $\pm$ 0.0029 & 0.6078 $\pm$ 0.0034 \\
DIEN  & 0.7712 $\pm$ 0.0033 & 0.7111 $\pm$ 0.0024 & 0.5915 $\pm$ 0.0027 \\
DSIN  & 0.7843 $\pm$ 0.0021 & 0.7246 $\pm$ 0.0017 & 0.5883 $\pm$ 0.0011 \\
BST   & 0.7978 $\pm$ 0.0014 & 0.7352 $\pm$ 0.0018 & 0.5626 $\pm$ 0.0016 \\
DMIN  & 0.8026 $\pm$ 0.0022 & 0.7398 $\pm$ 0.0014 & 0.5437 $\pm$ 0.0013 \\
HSTU  & 0.8239 $\pm$ 0.0019 & 0.7476 $\pm$ 0.0024 & 0.5118 $\pm$ 0.0015 \\
\textbf{Ours} & \textbf{0.8486 $\pm$ 0.0020} & \textbf{0.7612 $\pm$ 0.0012} & \textbf{0.4974 $\pm$ 0.0014} \\
\bottomrule
\end{tabular}
\end{table}

\subsection{Online Deployment}
GateDiffInt has been deployed in the ranking stage of the home feed of a major industrial platform serving hundreds of millions of daily active users. The system follows a two-stage paradigm: GMCD and MILD are pre-trained and updated weekly, while they are jointly fine-tuned with the CVR model via LoRA on a daily basis---the weekly cadence suffices because sequence enhancement and intent extraction capture relatively stable patterns, whereas the daily fine-tuning adapts to the latest data distribution. At serving time, the latest user behaviors are retrieved in real time and passed sequentially through GMCD, the MILD student, and the CVR head to produce the final scores. In a 14-day online A/B test against the production baseline, GateDiffInt achieves a statistically significant \textbf{GMV} improvement of \textbf{+1.13\%} in the home-feed scenario, and has since been rolled out to other businesses within the same scenario, bringing a cumulative \textbf{GMV} gain of \textbf{+5.13\%} to the feed scenario. These results confirm the practical value and deployability of GateDiffInt at industrial scale.

\subsection{Ablation Study}
To assess each component, we ablate GateDiffInt on Taobao (Table~\ref{tab:ablation}). Removing both GMCD and MILD causes the largest drop, and removing either alone also degrades performance---MILD more so. The joint drop is markedly larger than the sum of the two individual drops, indicating that the two modules are complementary rather than redundant. Within GMCD, replacing controllable diffusion with a vanilla variant, or disabling the fusion gate or importance gate, all hurt performance, with the fusion gate contributing most. Within MILD, collapsing the four per-intent LoRAs into one, or the four intent types into a single type, likewise degrades results. These confirm the necessity of gate-mediated control and multi-intent disentanglement.

\begin{table}[h]
\centering
\caption{Ablation study results in Taobao Dataset.}
\label{tab:ablation}
\begin{tabular}{lccc}
\toprule
Variant & AUC & GAUC & LogLoss \\
\midrule
\textbf{GateDiffInt} & \textbf{0.8515} & \textbf{0.8275} & \textbf{0.4836} \\
GateDiffInt w/  Vanilla Diffusion& 0.8326 & 0.8158 & 0.5112 \\
GateDiffInt w/o GMCD \& MILD     & 0.7154 & 0.6295 & 0.7128 \\
GateDiffInt w/o GMCD             & 0.8313 & 0.8139 & 0.5241 \\
GateDiffInt w/o MILD             & 0.8278 & 0.8037 & 0.5092 \\
GateDiffInt w/o Fusion Gate      & 0.8358 & 0.8170 & 0.5073 \\
GateDiffInt w/o Importance Gate  & 0.8424 & 0.8215 & 0.4983 \\
GateDiffInt w/o Multi-LoRA       & 0.8467 & 0.8253 & 0.4981 \\
GateDiffInt w/o Multi-Intent     & 0.8374 & 0.8133 & 0.5022 \\
\bottomrule
\end{tabular}
\end{table}

\section{Conclusion}
In this paper, we identify and empirically characterize \emph{Noise--Intent Coupling} (NIC) in CVR sequence modeling---the mutual reinforcement between noise-induced intent distortion (NID) and intent-agnostic denoising failure (IDF)---and propose GateDiffInt, which uses the supervisory signal of the final conversion task as a shared guide to align Gate-Mediated Controllable Diffusion (GMCD) with Multi-Intent LLM Distillation (MILD). Offline experiments on public and industrial datasets and a 14-day online A/B test show that GateDiffInt consistently improves ranking quality and calibration while delivering real business value. Future work will extend NIC to multi-task CVR/CTR modeling~\cite{39MMoE,40PLE} and further develop diagnostic characterizations of NID and IDF.

\bibliographystyle{ACM-Reference-Format}
\bibliography{reference}
\end{document}